\documentclass[final,1p,times]{elsarticle}

\usepackage
{
    amsmath,
    amssymb,
    amsthm,
    multicol,
    color
}

\theoremstyle{plain}
\newtheorem{lemma}{Lemma}[section]

\newtheorem{proposition}[lemma]{Proposition}

\theoremstyle{definition}
\newtheorem{definition}[lemma]{Definition}
\newtheorem{example}[lemma]{Example}
\newtheorem{remark}[lemma]{Remark}
\newtheorem{algorithm}[lemma]{Algorithm}

\providecommand{\norm}[1]{\lVert#1\rVert}
\providecommand{\coloneqq}{\mathrel{\mathop:}=}
\newcommand{\subfiguretitle}[1]{\scriptsize{#1} \\ \vspace*{1mm}}
\newcommand{\subs}[1]{\ensuremath{_{\textrm{#1}}}}
\newcommand{\mc}[1]{{\mathcal{#1}}}
\newcommand{\mf}[1]{{\mathfrak{#1}}}

\journal{Journal of Computational and Applied Mathematics}

\begin{document}

\begin{frontmatter}

\title{An efficient algorithm for the parallel solution of high-dimensional differential equations}

\author[a1]{Stefan Klus}
\author[a2]{Tuhin Sahai\corref{c1}}
\author[a2]{Cong Liu}
\author[a1]{Michael Dellnitz}

\address[a1]{Institute for Industrial Mathematics, University of Paderborn, 33095 Paderborn, Germany}
\address[a2]{United Technologies Research Center, East Hartford, CT 06108, USA}

\cortext[c1]{Corresponding author. Email: SahaiT@utrc.utc.com}

\begin{abstract}
The study of high-dimensional differential equations is
challenging and difficult due to the analytical and
computational intractability. Here, we improve the speed of
waveform relaxation (WR), a method to simulate high-dimensional
differential-algebraic equations. This new method termed
adaptive waveform relaxation (AWR) is tested on a communication
network example. Further we propose different heuristics for
computing graph partitions tailored to adaptive waveform
relaxation. We find that AWR coupled with appropriate graph
partitioning methods provides a speedup by a factor between $3$
and $16$.
\end{abstract}


\end{frontmatter}


\section{Introduction}
\label{sec:Introduction}

Over the past few years, several attempts have been made to
study differential equations of high dimensionality. These
equations naturally occur in models for systems as diverse as
metabolic networks \cite{MetaNet}, communication networks
\cite{McMillan}, fluid turbulence~\cite{Berkooz}, heart dynamics
\cite{heart}, chemical systems \cite{icepic} and electrical
circuits \cite{nevanlinear} to name but a few. Traditional
approaches approximate the full system by dynamical systems of
lower dimension. These model reduction techniques \cite{Modred}
include proper orthogonal decomposition (POD) along with
Galerkin projections \cite{Berkooz}, Krylov subspace methods
\cite{Krylov}, and balanced truncation or balanced POD (see e.g.
\cite{Row05}).

In this work, we accelerate a parallel algorithm, for the
simulation of differential-algebraic equations, called waveform
relaxation \cite{nevanlinear,Alberto,WaveformConv}. In waveform
relaxation, instead of approximating the original system by a
lower-dimensional model, the methodology is to distribute the
computations for the entire system on multiple processors. Each
processor solves only a part of the problem. The solutions
corresponding to subsystems on other processors are regarded as
inputs whose waveforms are given by the solution of the previous
iteration. This step is one iteration of the procedure. At the
end of each iteration the solutions are distributed among the
processors. The procedure is repeated until convergence is
achieved. The initial waveforms are typically chosen to be
constant.

This paper is organized as follows:  Based on previously derived
error bounds for waveform relaxation (cf. \cite{Bur95, BS97}),
we propose and demonstrate a new algorithm to break the time
interval for simulation $ [0, T] $ into smaller subintervals. We
call this method \emph{adaptive waveform relaxation}. It is
important to note that this method is different from windowing
methods discussed in~\cite{Alberto}. Subsequently, we analyze
and present time and memory complexity of waveform relaxation
techniques and the dependence of the convergence behavior on the
decomposition of the system. Furthermore, we introduce different
graph partitioning heuristics in order to efficiently generate
an appropriate splitting. We demonstrate that the combination of
graph partitioning along with adaptive waveform relaxation
results in an improved performance over traditional waveform
relaxation and standard windowing techniques.

\section{Error bounds}

For an ordinary differential equation of the form $ \dot{x} =
f(x) $, $ f : \mathbb{R}^n \mapsto \mathbb{R}^n $, the iteration
method described in the introduction can be written as
\begin{equation} \label{nonlinwr}
    \dot{x}^{k+1} = \phi(x^{k+1}, x^{k}),
\end{equation}
with $ \phi : \mathbb{R}^n \times \mathbb{R}^n \mapsto
\mathbb{R}^{n} $ and $ \phi(x, x) = f(x) $. The standard
Picard--Lindel\"of iteration, for example, is given by $ \phi(x,
y) = f(y) $. Convergence is, by definition, achieved if $
\norm{x^{k+1} - x^k} < \varepsilon $ for a predefined threshold
$ \varepsilon $. This procedure can be used to solve
differential-algebraic equations as well. For a more detailed
overview on waveform relaxation we refer
to~\cite{nevanlinear,Alberto}. We assume that the splitting $
\phi $ is Lipschitz continuous, i.e.\ there exist constants $
\mu \ge 0 $ and $ \eta \ge 0 $ such that
\begin{equation} \label{Lipschitz}
    \norm{\phi(x, y) - \phi(\tilde{x}, \tilde{y})} \le \mu \norm{x - \tilde{x}} + \eta \norm{y - \tilde{y}}.
\end{equation}
Let $ \bar{x} $ be the exact solution of the differential
equation and define $ E_k $ to be the error of the $ k $-th
iterate, that is
\begin{equation}
    E_k = x^k - \bar{x}.
\end{equation}
It is well known that the iteration given by Eqn.~\ref{nonlinwr}
converges superlinearly (evident in Proposition $2.1$) to the
exact solution and that the error is bounded. Convergence
results and error bounds for waveform relaxation have previously
been derived in \cite{Alberto,WaveformConv,Bur95,BS97}. For the
purpose of this paper the following version of the convergence
result will be useful.

\begin{proposition}
Assuming that the splitting $ \phi $ satisfies the Lipschitz
condition, the norm of the error $ \norm{E_{k}} $ on the
interval $ [0, T] $ is bounded as follows
\begin{equation} \label{BoundsTheo}
    \norm{E_k} \leq \frac{C^k \eta^k T^k}{k!} \norm{E_0},
\end{equation}
with $ C = e^{\mu T} $.
\end{proposition}

\begin{remark}
In Eqn.~\ref{BoundsTheo} it is important to note that $ k! $
will eventually dominate the numerator such that convergence is
guaranteed.
\end{remark}

\section{Adaptive waveform relaxation}

By Eqn.~\ref{BoundsTheo} the error of standard waveform
relaxation crucially depends on $ T $. The longer the time
interval, the greater is the number of iterations needed to
bound the error below a desired tolerance. This fact is well
known and in \cite{Alberto} it is suggested to subdivide the
time interval $ [0, T] $ into \textit{windows} $ [0, T_1] $, $
[T_1, T_2] $, $ \dots $, $ [T_{\nu-1}, T_{\nu}] $. The authors
pick an initial interval of $ \frac{T}{20} $ and then perform
waveform relaxation on the small interval. If the solution has
not converged in $ 5 $ iterations, then the time window is
halved. If the size of the interval is too large (based on data
storage requirements), the window length is reduced. If the
current window satisfies the above requirements, the same window
length is used for the next interval. This approach does not
take into account the slope of the solution and the error made
by the initial waveform. We aim to adaptively determine the size
of the next time interval based on the previously computed
solution and on Eqn.~\ref{BoundsTheo}.

Let us be more precise. In our procedure, we too first perform waveform relaxation on a small interval given by $ [0, T_1] $. Define $ \Delta T_i = T_i - T_{i-1} $. Upon convergence of waveform relaxation on the interval $ [T_{i-1}, T_i] $, we estimate the length of the next time interval $ \Delta T_{i+1} $ as follows: Firstly, we compute an interpolating polynomial of order $ l $ using $ l + 1 $ equally spaced points $ t_j $, $ j = 0, \dots, l $. In our implementation, a quadratic polynomial with $ t_0 = T_i $, $ t_1 = T_i - \frac{1}{10} \Delta T_i $, and $ t_2 = T_i - \frac{2}{10} \Delta T_i $ is used. This interpolating polynomial is also utilized as an initial guess for the waveform over the next time interval. Using Eqn.~\ref{BoundsTheo}, we then choose $ \Delta T_{i+1} $ such that
\begin{equation} \label{Test}
    \norm{\hat{E}_{i+1, r}} \coloneqq \frac{ \left(e^{\mu \Delta T_{i+1}} \eta \Delta T_{i+1}\right)^r}{r!}
        \norm{E_{i+1, 0}} < \varepsilon.
\end{equation}
In other words, given a desired number of iterations $ r $, one
can estimate the length of the next time interval if $
\norm{E_{i+1, 0}} $, $ \mu $, and $ \eta $ are known. To
estimate the error $ E_{i+1, 0}(t) $, we compute the difference
between $ x^{k+1}(t) $ and the interpolating polynomial. This
can be accomplished using the formula
\begin{equation} \label{E0form}
    \tilde{E}_{i+1, 0}(t) = \frac{\phi^{(l)}(x^{k+1}(\xi), x^{k+1}(\xi))}{(l+1)!} \, \omega(t),
\end{equation}
where
\begin{equation} \label{omegaform}
    \omega(t) = (t - t_0)(t - t_1) \hdots (t - t_l)
\end{equation}
and $ \phi^{(l)} = \frac{d^l}{dt^l} \phi $ is the $ l $-th
derivative of the splitting $ \phi $ with respect to $ t $
(cf.~\cite{Stoer}). Additionally, we assume that $\phi^{(l)}$ in
the above equation exists. We estimate the magnitude of this
term using finite differences at the end of the time interval
just computed. The Lipschitz constants $ \mu $ and $ \eta $ also
need to be estimated in order to get a good guess for the
interval length. For nonlinear problems, the Lipschitz constants
are in general not directly available. Below, we will focus on
linear ordinary differential equations so that the Lipschitz
constants are given by the norms of the matrix splitting, as we
will show in Section~\ref{sec:Partitioning and Convergence}.

With an estimate of all the variables in Eqn.~\ref{Test} we can
now compute the length of the next window. Initially, we set $
\Delta T_{i+1} = 2 \Delta T_i $ and compute $ \tilde{E}_{i+1,
0}(T_i + \Delta T_{i+1}) $. This gives an estimate for the
magnitude of $ \norm{E_{i+1, 0}} $ for the next time interval.
If the resulting error $ \norm{\hat{E}_{i+1, r}} $ is larger
than the threshold $ \varepsilon $, we repeat the process using
an adapted interval length $ \Delta T_{i+1} $ as described in
the following algorithm.

\begin{algorithm} \label{alg:AWR}
To compute the length $ \Delta T_{i+1} $, execute the following steps:
\begin{enumerate}
\item Set $ \Delta T_{i+1} = 2 \Delta T_i $ and $ \delta = \frac{1}{20} \Delta T_i $.
\item Evaluate $ \tilde{E}_{i+1, 0}(T_i + \Delta T_{i+1}) $ using Eqn.~\ref{E0form} to estimate $ \norm{E_{i+1, 0}} $ and compute $ \norm{\hat{E}_{i+1, r}} $ with the aid of Eqn.~\ref{Test}.
\item If $ \norm{\hat{E}_{i+1, r}} > \varepsilon $ and $ \Delta T_{i+1} > \frac{1}{2} \Delta T_i $, set $ \Delta T_{i+1} = \Delta T_{i+1} - \delta $ and repeat step~2.
\end{enumerate}
\end{algorithm}

We define the minimal window length to be $ \Delta T_{i+1} =
\frac{1}{50} T $. The above procedure gives a sequence of time
intervals $ [0, T_1] $, $ [T_1, T_2] $, $ \dots $, $ [T_{\nu-1},
T_{\nu}] $, where $ T_{\nu} = T $,  on which waveform relaxation
is performed with an initial ``guess'' waveform provided by an
extrapolation of the solution on the previous interval.

Intuitively, this procedure works by taking small steps in
regions where the solution changes rapidly (large derivative)
and large steps in regions where the solution changes slowly
(small derivative).

\section{Partitioning and convergence}
\label{sec:Partitioning and Convergence}

In this section, we analyze the time and memory complexity of waveform relaxation and the influence of the splitting on the convergence. It is shown that the optimal splitting depends on both the integration scheme and the step size. Since there exists no efficient method to compute the optimal splitting directly, we introduce different heuristics in order to generate appropriate decompositions. Here, we focus on linear systems of the form
\begin{equation} \label{Fulleqns}
    \dot{x}(t) = Q x(t),
\end{equation}
with $ Q \in \mathbb{R}^{n \times n} $, $ x \in \mathbb{R}^n $, $ t \in [0, \tau] $, and the initial condition $ x(0) = x_0 $. Linear equations arise in models of various dynamical systems. We will consider in particular systems which are derived from generalized stochastic Petri nets. In order to solve the initial value problem with the aid of waveform relaxation or adaptive waveform relaxation, the system is split according to $ P Q P^T = M + N $ and the partitioned system
\begin{equation} \label{eqnwr}
    \dot{x}^{k+1}(t) = M x^{k+1}(t) + N x^k(t)
\end{equation}
is solved iteratively. Here, $ P $ is a permutation matrix and $ M $ is a block diagonal matrix. Hence, $ \phi(x^{k+1}, x^k) = M x^{k+1} + N x^k $. Furthermore, the Lipschitz constants $ \mu $ and $ \eta $ are the appropriate matrix norms of $ M $ and $ N $, respectively. The matrix splitting can be regarded as a graph partitioning problem where each block of $ M $ represents a part or subsystem and $ N $ the connections between different parts. Let $ p $ be the number of blocks where the $ i $-th block is of size $ n_i $, that is $ n = \sum_{i=1}^p n_{i} $. Then the $ i $-th equation can be written as
\begin{equation} \label{ieqnwr}
    \dot x_i^{k+1}(t) = M_{ii} x_{i}^{k+1}(t) + \sum_{j\neq i} N_{ij} x_{j}^k(t),
\end{equation}
with $ M_{ii} \in \mathbb{R}^{n_i \times n_i} $, $ x_i \in \mathbb{R}^{n_i} $, $ N_{ij} \in \mathbb{R}^{n_{i}\times n_{j}} $, and $ x_j \in \mathbb{R}^{n_j} $ for $ j = 1, \dots, p $ and $ j \ne i $.

Let us begin with a remark on the time and memory complexity of waveform relaxation. Our aim is to derive conditions under which one expects waveform relaxation (in a parallel implementation) to give an answer faster than solving the entire system of equations (in a serial implementation). For simplicity, we consider the explicit Euler method with a fixed step size $ h $. The same argument can be repeated for other integration schemes with the same result.

Elementary calculations show that for the full system \eqref{Fulleqns} the cost of the numerical solution on the interval $ [0, \tau] $ amounts to
\begin{equation}
    C_\text{E} = (n^2 + n) \frac{\tau}{h}.
\end{equation}
We now compute the time complexity of waveform relaxation. The cost of a single Euler step for the $ i $-th subsystem \eqref{ieqnwr} is $ C_{\text{WR}_i} = n_i^2 + n_i (n - n_i) + n_i $. Thus, to compute $ K $ iterations for all blocks, the total cost would be
\begin{equation}
    C_\text{WR} = K(n^2 + n) \frac{\tau}{h} = K C_\text{E}.
\end{equation}
Let us assume that there are $ p $ processors, and let the $ l $-th block be the largest, then the time complexity in the parallel case is given by
\begin{equation}
    C_\text{WRp} = K (n_l n + n_l) \frac{\tau}{h}.
\end{equation}
It follows that if $ n_l K < n $, then the waveform relaxation procedure is advantageous. Note that $ K $, or the number of iterations needed for convergence, strongly depends on the actual decomposition.

The memory complexity in the linear case is easy to classify. In general, one needs to store a big matrix of size $ n^2 $. On a single processor, waveform relaxation has the same memory requirements as the full system. For the parallel case, however, the maximum storage needed is $ n_l \times n $. This can be a major advantage
if $ n_l \ll n $ and the matrix can be stored in the processor cache. It is also important to note that the above analysis does not take communication costs into account.

\begin{remark}
In a nutshell, standard waveform relaxation is of advantage if
\begin{enumerate}[i)]
\item $ n_l K < n $ for time complexity,
\item $ n_l \ll n $ for memory complexity,
\end{enumerate}
where $ n_l $ is the size of the largest block of the decomposed system and $ K $ is the number of iterations needed for convergence.
\end{remark}

Let us now analyze the influence of the decomposition on the convergence. We discretize the system \eqref{eqnwr} using a fixed step size $ h $ and an integration scheme of the form
\begin{equation} \label{IntScheme}
    x^{k+1}_{m+1} = C_1 x^{k+1}_m + C_2 x^k_m + C_3 x^k_{m+1},
\end{equation}
where $ C_1 $, $ C_2 $, and $ C_3 $ are matrices which may depend on $ M $, $ N $, and $ h $. Let $ s = \frac{\tau}{h} $ be the number of time steps and $ X^k = [x^k_1 \; x^k_2 \; \dots \; x^k_s] $ the discretized waveform. Furthermore, define
\begin{equation}
    \hat{X}^k =
    \begin{bmatrix}
        x^k_1 \\
        x^k_2 \\
        \vdots \\
        x^k_s
    \end{bmatrix}.
\end{equation}

\begin{proposition}
For an integration scheme of the form \eqref{IntScheme} the discrete waveform relaxation can be written as $ \hat{X}^{k+1} = A \hat{X}^k + b $, with
\begin{equation}
    A =
    \begin{bmatrix}
        C_3                             \\
        C_1 C_3 + C_2 & C_3                   \\
        C_1^2 C_3 + C_1 C_2 & C_1 C_3 + C_2 & C_3         \\
        \vdots & \ddots & \ddots & \ddots \\
        C_1^{s-1} C_3 + C_1^{s-2} C_2 & \dots & C_1^2 C_3 + C_1 C_2 & C_1 C_3 + C_2 & C_3
    \end{bmatrix}
\end{equation}
and
\begin{equation}
    b =
    \begin{bmatrix}
        (C_1 + C_2) \, x_0 \\
        C_1 (C_1 + C_2) \, x_0 \\
        C_1^2 (C_1 + C_2) \, x_0 \\
        \vdots \\
        C_1^{s-1} (C_1 + C_2) \, x_0
    \end{bmatrix}.
\end{equation}
\end{proposition}

\begin{proof}
By Eqn. \ref{IntScheme}
\begin{equation*}
    \hat{X}^{k+1} =
    \underbrace{
    \begin{bmatrix}
        0   &                 \\
        C_1 & 0               \\
            & \ddots & \ddots \\
            & & C_1 & 0
    \end{bmatrix}}_{\displaystyle U}
    \hat{X}^{k+1} +
    \underbrace{
    \begin{bmatrix}
        C_3 &                 \\
        C_2 & C_3             \\
            & \ddots & \ddots \\
            & & C_2 & C_3
    \end{bmatrix}}_{\displaystyle V}
    \hat{X}^k +
    \underbrace{
    \begin{bmatrix}
        (C_1 + C_2) \, x_0 \\
        0 \\
        \vdots \\
        0
    \end{bmatrix}}_{\displaystyle d}
\end{equation*}
and thus $ \hat{X}^{k+1} = (I - U)^{-1} V \hat{X}^k + (I - U)^{-1} d $, where $ I $ is the identity matrix. Using the Neumann series and the fact that $ U $ is nilpotent, we get
\begin{equation*}
    \begin{split}
        (I - U)^{-1} &= \sum_{i=0}^{s-1} U^i
           = \begin{bmatrix}
                 I       &                         \\
                 C_1     & I                       \\
                 C_1^2   & C_1   & I               \\
                 \vdots & \ddots & \ddots & \ddots \\
                 C_1^{s-1} & \dots & C_1^2 & C_1 & I
             \end{bmatrix}.
    \end{split}
\end{equation*}
Hence, $ A = (I - U)^{-1} V $ and $ b = (I - U)^{-1} d $ are of the aforementioned form.
\end{proof}

\begin{example}
The following integration schemes are of the form \eqref{IntScheme}:
\begin{enumerate}[i)]
\item Explicit Euler method: $ C_1 = (I + h M) $, $ C_2 = h N $, and $ C_3 = 0 $.
\item Implicit Euler method: $ C_1 = (I - h M)^{-1} $, $ C_2 = 0 $, and $ C_3 = (I - h M)^{-1} h N $.
\item Trapezoidal rule: $ C_1 = (I - \frac{h}{2} M)^{-1} (I + \frac{h}{2} M) $ and $ C_2 = C_3 = (I - \frac{h}{2} M)^{-1} \frac{h}{2} N $.
\end{enumerate}
\end{example}

To begin with, we discretize the system using the explicit Euler method. Since $ C_3 = 0 $, $ A $ is a strictly lower-triangular block Toeplitz matrix. It follows that the spectral radius $ \rho(A) $ is $ 0 $ and in particular $ A^s = 0 $. Therefore, waveform relaxation converges, independent of the partitioning, after at most $ s + 1 $ iterations, i.e.
\begin{equation}
    \begin{split}
        \hat{X}^s &= A \hat{X}^{s-1} + b = \underbrace{A^s \hat{X}^0}_{0} + A^{s-1} b + \dots + A b + b\,, \\
        \hat{X}^{s+1} &= A \hat{X}^s + b = \underbrace{A^s b}_{0} + A^{s-1} b + \dots + A b + b = \hat{X}^s.
    \end{split}
\end{equation}

If we replace the explicit Euler method by the implicit Euler method, then the spectral radius of $ A $ is equal to the spectral radius of $ C_3 = (I - h M)^{-1} h N $. To accelerate the convergence of waveform relaxation, the matrix $ Q $ should be decomposed such that the spectral radius of $ C_3 $ is minimized. Observe that the optimal splitting depends on the step size $ h $.

If we, on the other hand, use the trapezoidal rule, then the block diagonal of the iteration matrix $ A $ is given by $ C_3 = (I - \frac{h}{2} M)^{-1} \frac{h}{2} N $. That is, the system should be partitioned in a way that the spectral radius of the new matrix $ C_3 $ is minimized. Thus, the optimal splitting depends also on the integration scheme.

Since the iteration matrices $ A $ of the implicit Euler or the trapezoidal rule based waveform relaxation are highly nonnormal, their spectral properties do not predict the convergence behavior appropriately. For such matrices and operators the pseudospectrum is a more useful tool~\cite{LW97}.

\begin{definition}
Given a matrix $ A $ and $ \varepsilon > 0 $, $ \lambda \in \mathbb{C} $ is defined to be an $ \varepsilon
$-\emph{pseudoeigenvalue} of $ A $ if $ \lambda $ is an eigenvalue of $ A + E $ for a matrix $ E $ with $ \norm{E} < \varepsilon $.
\end{definition}

There are several different equivalent definitions of pseudo-eigenvalues (cf.~\cite{TE05}). The set $ \Lambda_\varepsilon(A) $ of all $ \varepsilon $-pseudoeigenvalues is called the $ \varepsilon $-pseudo\-spectrum and $ \rho_\varepsilon(A) = \max\{ \norm{z} \mid z \in \Lambda_\varepsilon(A) \} $ is called the $ \varepsilon $-pseudospectral radius. While the $ \varepsilon $-pseudo\-spectrum of a normal matrix is the union of $ \varepsilon $-balls around the eigenvalues, the pseudospectrum of a nonnormal matrix can be sensitive to small perturbations~\cite{JO95}.

In Section~\ref{sec:Applications and Results}, the matrix splittings with the best spectral and pseudospectral properties are used for comparison. However, there exists no efficient method to minimize the spectral radius or the pseudospectral radius directly. We propose different heuristics to find a decomposition which is close to the optimal splitting. The partitioning of a directed graph with respect to a given cost function is still an open problem, in particular there are no sophisticated spectral clustering methods for directed graphs (cf.~\cite{Hen03}). Therefore, we combine different graph clustering and partitioning methods, namely horizontal-vertical decomposition, spectral clustering, and the graph partitioning library PARTY, to generate appropriate splittings.

\emph{Horizontal-vertical decomposition} as described in \cite{VKLM04} identifies the subsystem hierarchy of dynamical systems. The decomposition is equivalent to the computation of the strongly connected components of the graph $ \mf{G}(Q) $, where $ \mf{G}(Q) = (\mf{V}, \mf{E}) $ with $ \mf{V} = \{ \mf{v}_1, \dots, \mf{v}_n \} $ and $ \mf{E} = \{ (\mf{v}_i, \mf{v}_j) \mid q_{ij} \ne 0 \} $. The strongly connected components can be computed efficiently using the depth-first search.

\emph{Spectral Clustering} is a popular partitioning heuristic for undirected graphs, based on spectral or algebraic graph theory. Spectral clustering utilizes the information obtained from eigenvalues and eigenvectors of graph-related matrices such as the graph Laplacian for partitioning. For a detailed description we refer to \cite{Tutorial}. Recently, an efficient distributed spectral clustering algorithm that overcomes the drawbacks associated with random walk based approaches has been proposed by one of the authors in \cite{SSB09}.

\emph{PARTY} is a graph partitioning library that provides several different multilevel graph partitioning strategies combining local and global heuristics for undirected graphs \cite{Pre00}. The idea of the multilevel approach is to coarsen the initial graph by collapsing matching vertices so that global partitioning heuristics can be applied efficiently. Subsequently, combined vertices are split during the refinement process and local methods like the Kernighan--Lin heuristic or the Helpful-Set algorithm are applied to further improve the partition.

If the matrix $ Q $ is reducible, then the system is decomposed first using the horizontal-vertical decomposition in order to exploit the directionality of the graph on a coarse level. Then, depending on the application, either the spectral clustering method or PARTY is applied to the individual strongly connected components. Since both methods are confined to undirected graphs, the strongly connected components have to be regularized first by omitting the orientation of the edges. If it is important to generate a balanced partition of the graph, then PARTY is, in general, better suited. If, on the other hand, the network is quite inhomogeneous and the spectral method computes an unbalanced splitting while PARTY is forced to generate a balanced splitting, then spectral partitioning is advantageous.

For large networks with several strongly connected components, the hori\-zontal-vertical decomposition is crucial for the quality of the decomposition. If the partitioning methods are directly applied to the graph $ \mf{G}(Q + Q^T) $, all information on the directed signal flow and the different subsystems is lost. In the next section we will demonstrate the impact of the horizontal-vertical decomposition on the convergence of waveform relaxation.

\section{Applications and results}
\label{sec:Applications and Results}

To illustrate the adaptive waveform relaxation procedure and the spectral and pseudospectral properties of the iteration matrices, we analyze a linear ordinary differential equation that is used for the transient analysis of a continuous-time Markov chain (CTMC). The continuous-time Markov chain is derived from a generalized stochastic Petri net (GSPN) \cite{GSPNBook}. GSPN is a popular model for performance analysis of complex concurrent systems. It has been used to model and analyze communication protocols \cite{GSPNATM}, parallel programs \cite{GSPNSoft}, multiprocessor architectures \cite{GSPNHard}, and manufacturing systems \cite{GSPNMaintenance}. The \emph{reachability graph} of a GSPN with an initial marking (state) consists of vertices corresponding to its reachable markings and directed edges corresponding to transitions. It has been proved that there exists a one-to-one mapping between the reachability graph of a GSPN and the CTMC \cite{GSPNequalCTMC}.

\begin{figure}[htb]
    \centering
    \includegraphics[width=0.3\textwidth]{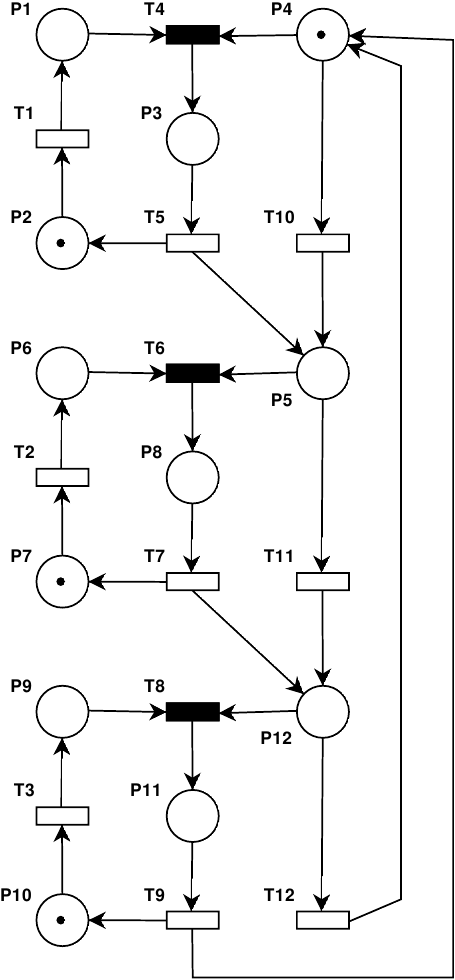}
    \caption{A GSPN model of a client-server system.}
    \label{GSPNFigure}
\end{figure}

Let $ \pi_i(t) $ be the probability that the CTMC is in state $ i $ at time $ t $. Let $ r_{ij} $, $ i \neq j $, be the transition rate from state $ i $ to state $ j $ and $ r_{ii} = -\sum_{j \neq i} r_{ij} $. Given the \emph{transition rate matrix} $ R = [r_{ij}] $ of a CTMC, the state probability distribution at time $ t $ denoted by $ \pi(t) = [\pi_1(t), \pi_2(t), \dots, \pi_n(t)] $ satisfies
\begin{equation} \label{CTMCeqnSys}
    \dot{\pi}(t) = \pi(t) R.
\end{equation}
We apply the waveform relaxation techniques to solve the above equation. Due to state space explosion, the number of differential equations becomes extremely large even for a GSPN of moderate size \cite{GSPNLarge}. This makes them an ideal application to demonstrate the waveform relaxation procedure. Figure~\ref{GSPNFigure} shows the GSPN that we used for experiment. It models a server shared by three clients. The corresponding CTMC of the GSPN has $ 24 $ states and the resulting transition rate matrix $ R $ is sparse.

For simplicity, we rewrite Eqn.~\ref{CTMCeqnSys} as $ \dot{x}(t) = Q x(t) $. In order to demonstrate the adaptive waveform relaxation procedure and to compare it to standard waveform relaxation, we decompose the GSPN into two subsystems of the same size.

\begin{figure}[htb]
    \centering
    \begin{minipage}[c]{0.49\textwidth}
        \centering
        \subfiguretitle{a) waveform relaxation}
        \includegraphics[width=\textwidth]{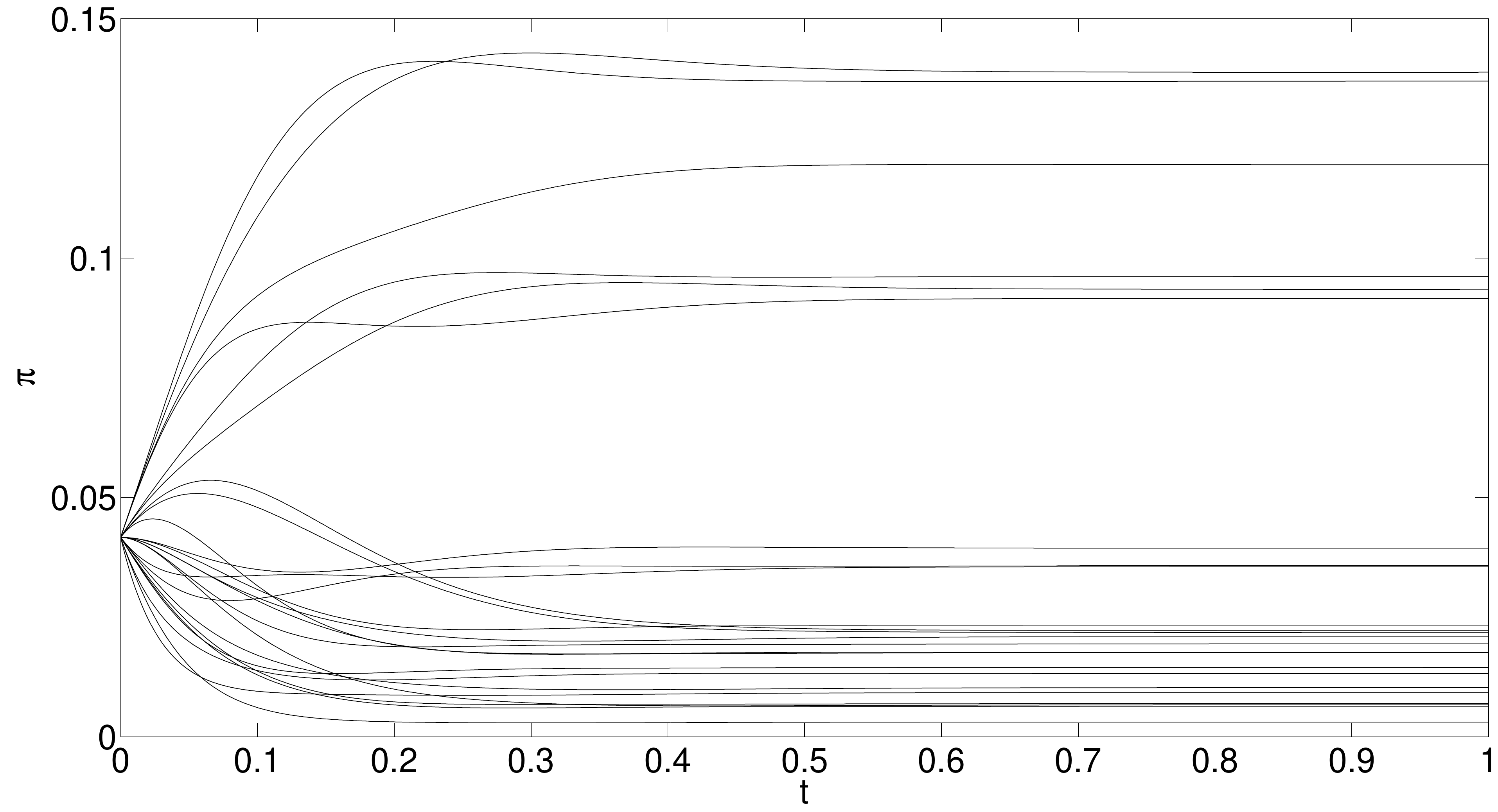}
    \end{minipage}
    \begin{minipage}[c]{0.49\textwidth}
        \centering
        \subfiguretitle{b) adaptive waveform relaxation}
        \includegraphics[width=\textwidth]{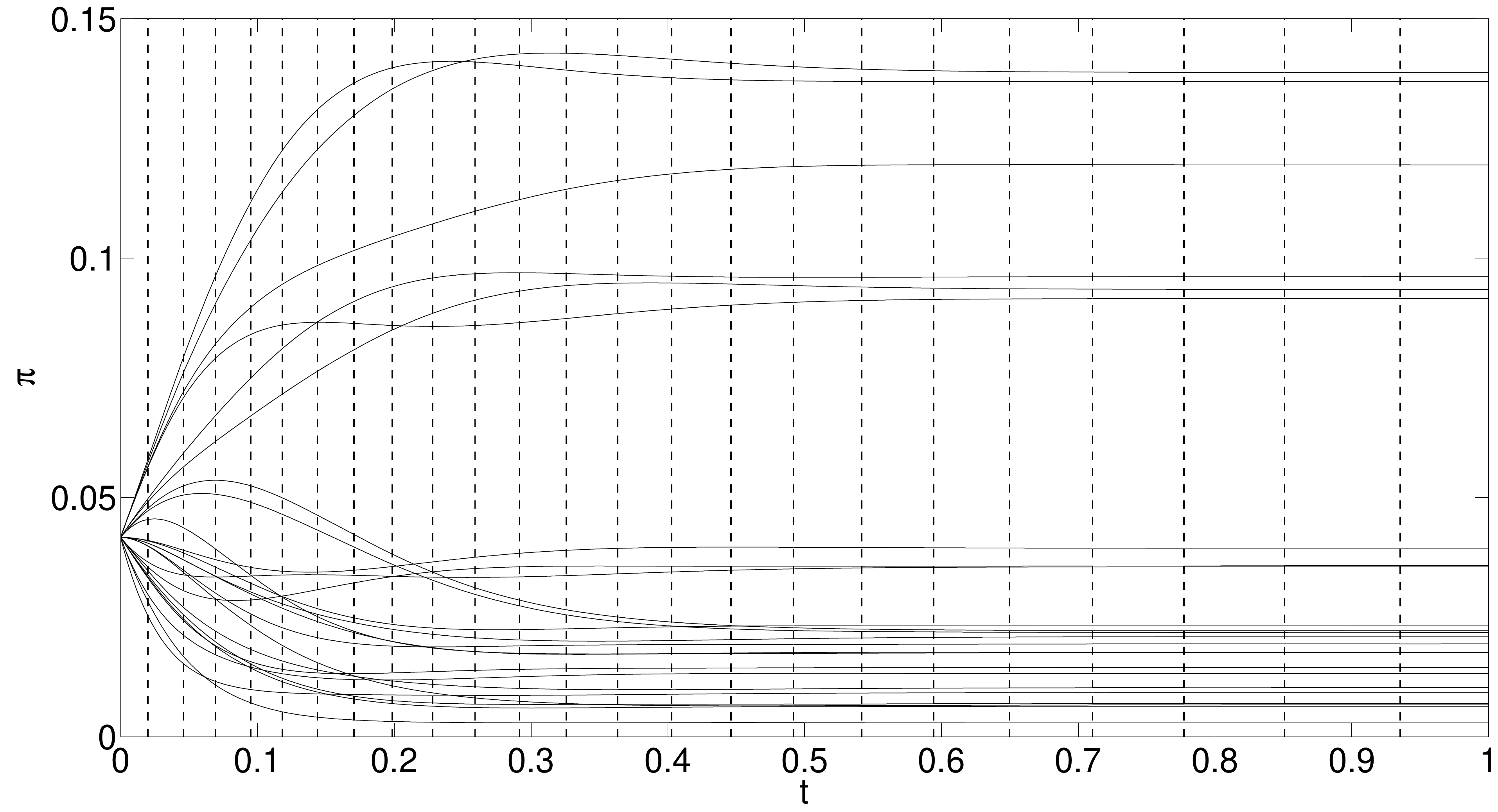}
    \end{minipage}
    \caption{Comparison of solutions obtained from waveform relaxation and adaptive waveform relaxation.}
    \label{fig:WR_AWR}
\end{figure}

Firstly, we compute the solution of the system using standard waveform relaxation and a fixed step size $ h = 10^{-3} $. The initial waveform is assumed to be constant over $ [0, T] $, i.e.\ $ x^0(t) = x_0 $, where $ T = 1 $. We iterate until the difference between two successive iterations falls below the predefined tolerance $ \varepsilon = 10^{-4} $. The solution is shown in Figure~\ref{fig:WR_AWR}a. As one can see, the state probability distributions approach constants, i.e.\ equilibria are eventually reached. Standard waveform relaxation takes (averaged over $ 10 $ simulations) $ 0.622 $ sec.

The solution is now computed using adaptive waveform relaxation. We use the same tolerance of $ \varepsilon = 10^{-4} $ and an initial window of $ [0, \frac{T}{50}] $. The solution and the intervals computed by adaptive waveform relaxation are shown in Figure~\ref{fig:WR_AWR}b. Averaging again over $ 10 $ simulations, we find that adaptive waveform relaxation takes approximately $ 0.103 $ sec to compute the solution, i.e.\ over $ 6 $ times faster than standard waveform relaxation.

\begin{figure}[htb]
    \centering
    \begin{minipage}[c]{0.45\textwidth}
        \centering
        \subfiguretitle{a) $ h = 10^{-1} $, $ \rho = 0.23632 $}
        \includegraphics[width=0.9\textwidth]{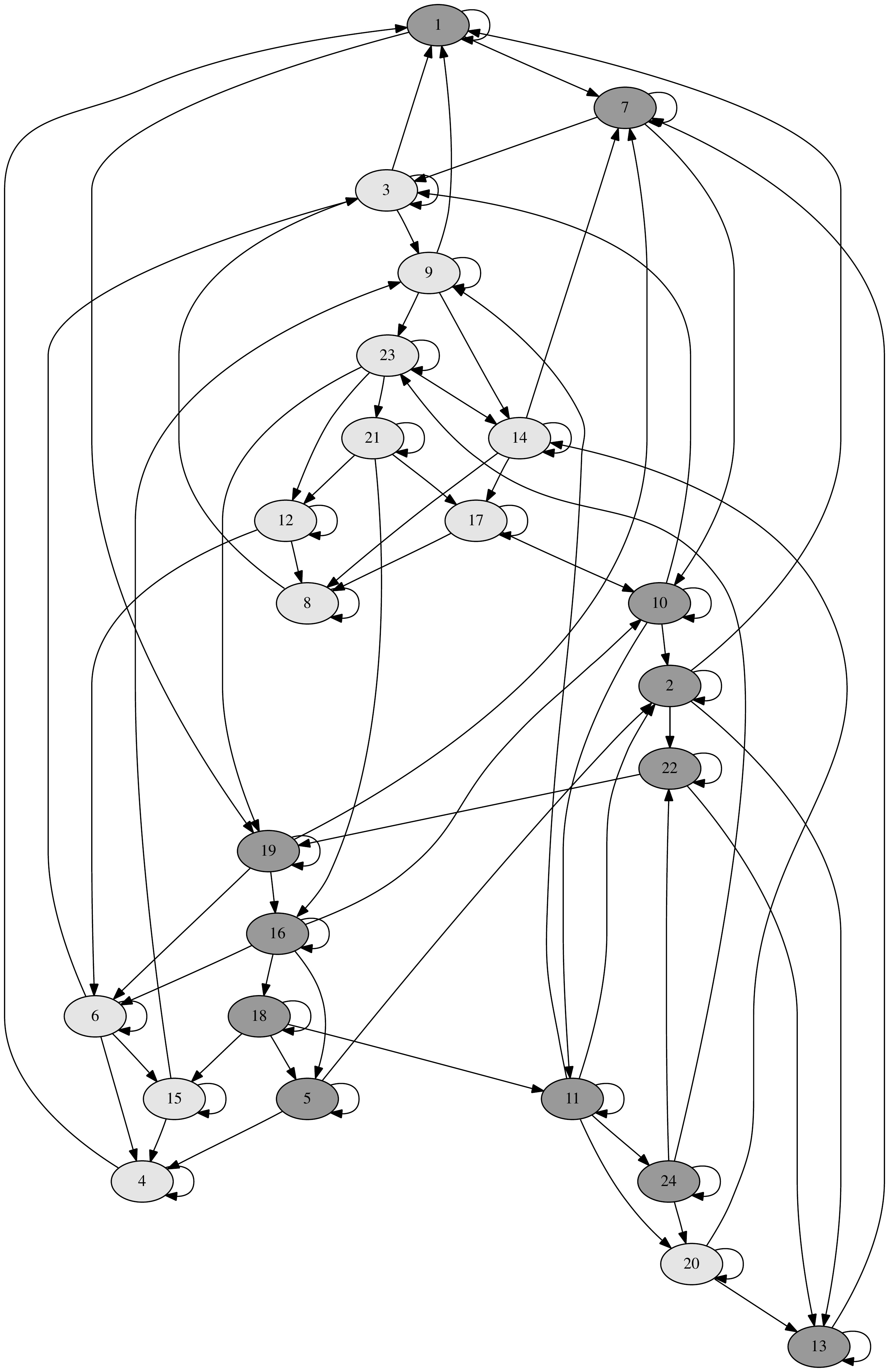}
    \end{minipage}
    \begin{minipage}[c]{0.45\textwidth}
        \centering
        \subfiguretitle{b) $ h = 10^{-2} $, $ \rho = 0.00505 $}
        \includegraphics[width=0.9\textwidth]{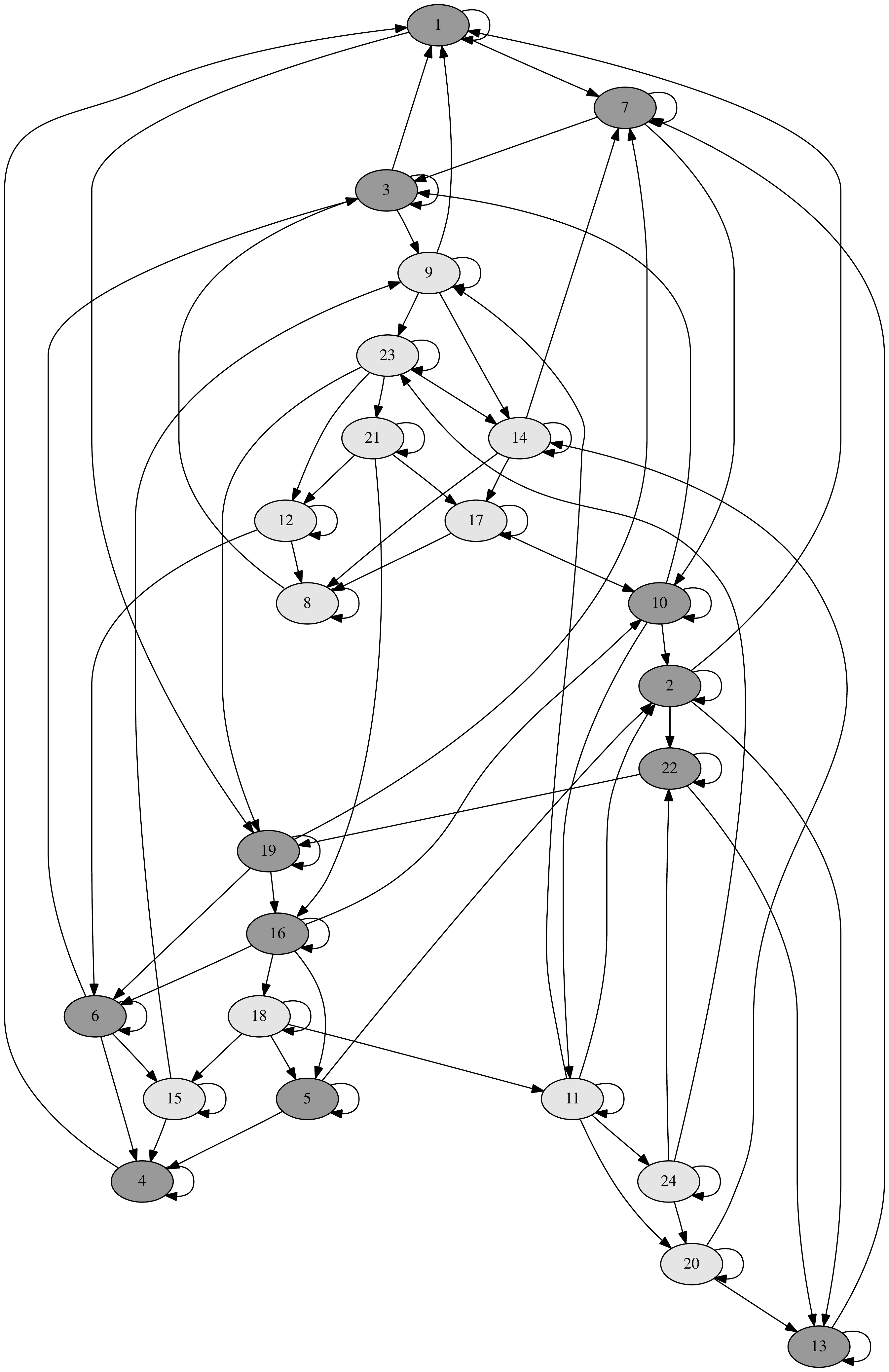}
    \end{minipage}
    \caption{Optimal splittings $ \mc{P}_1 $ and $ \mc{P}_2 $.}
    \label{fig:OptPart}
\end{figure}

In the following, we analyze different partitions of the GSPN to illustrate the influence of the matrix splitting on the convergence of waveform relaxation. The best balanced bipartition of the GSPN for the implicit Euler based waveform relaxation and $ h = 10^{-1} $ is given by
\begin{equation}
    \mc{P}_1 =
        [1 \; 2 \; 5\; 7 \; 10 \; 11 \; 13 \; 16 \; 18 \; 19 \; 22 \; 24 \mid
         3 \; 4 \; 6 \; 8 \; 9 \; 12 \; 14 \; 15 \; 17 \; 20 \; 21 \; 23]\,,
\end{equation}
meaning that the first 12 states belong to the first and the remaining 12 states to the second part, whereas for $ h = 10^{-2} $ the bipartition with the lowest spectral radius is
\begin{equation}
    \mc{P}_2 =
        [1 \; 2 \; 3 \; 4 \; 5 \; 6 \; 7 \; 10 \; 13 \; 16 \; 19 \; 22 \mid
         8 \; 9 \; 11 \; 12 \; 14 \; 15 \; 17 \; 18 \; 20 \; 21 \; 23 \; 24]\,.
\end{equation}
The splittings $ \mc{P}_1 $ and $ \mc{P}_2 $ are shown in Figure~\ref{fig:OptPart}a and \ref{fig:OptPart}b, respectively. If we use the trapezoidal rule, then for $ h = 10^{-1} $ and $ h = 10^{-2} $ the optimal splittings are again given by $ \mc{P}_1 $ and $ \mc{P}_2 $. Nevertheless, for $ h = 5 \cdot 10^{-2} $, for instance, $ \mc{P}_1 $ is better suited for the implicit Euler based waveform relaxation while $ \mc{P}_2 $ is better suited for the trapezoidal rule based waveform relaxation. This example illustrates that the optimal splitting depends on the step size and on the integration scheme. To compute these optimal partitions, we compared all balanced decompositions of the network. For high-dimensional systems this is clearly not feasible.

\begin{figure}[htb]
    \centering
    \includegraphics[width=0.55\textwidth]{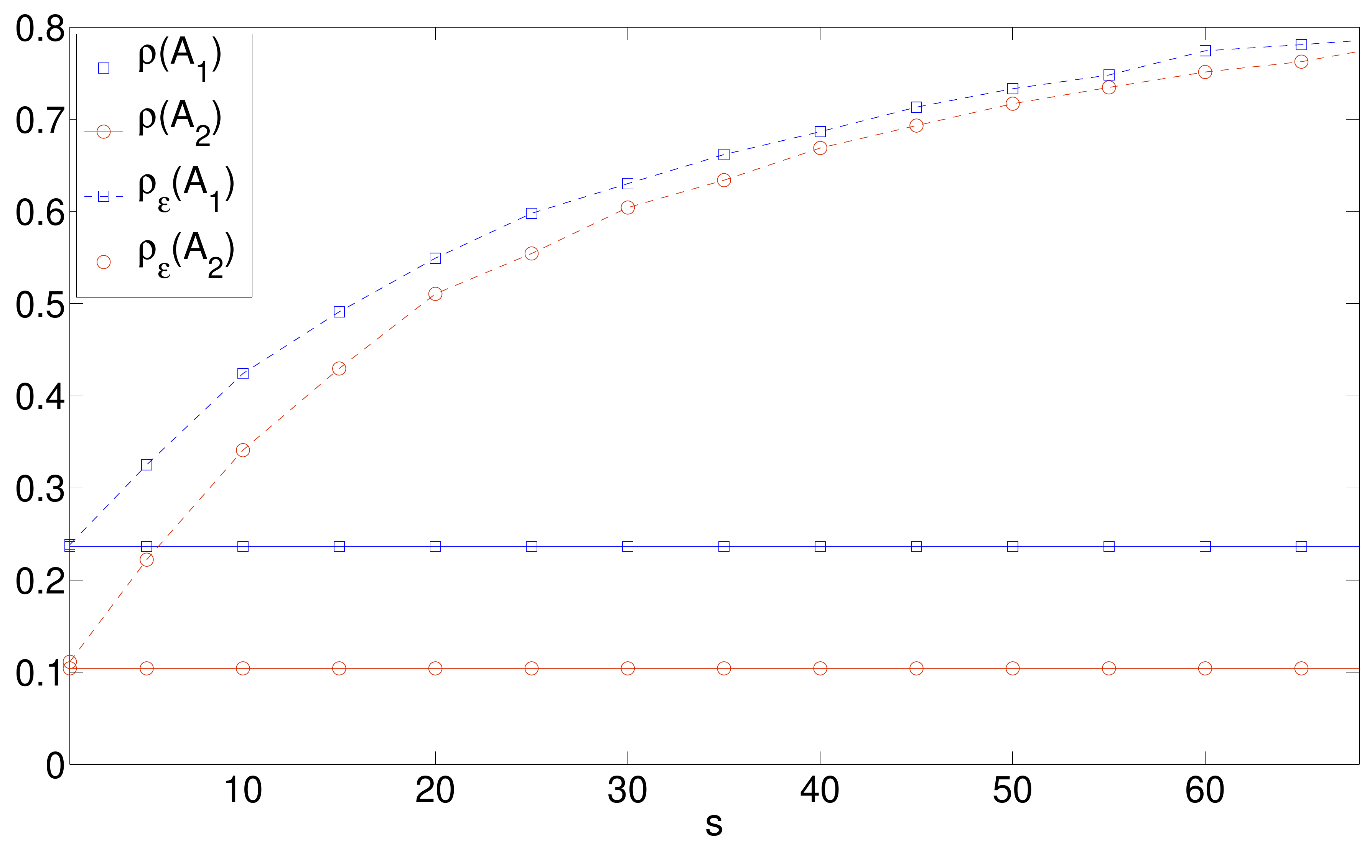}
    \caption{Dependence of the pseudospectral radii on the number of time steps.}
    \label{fig:sr_psr}
\end{figure}

From now on, we denote the waveform relaxation operator of the implicit Euler based method as $ A_1 $ and the operator of the trapezoidal rule based method as $ A_2 $. Although the spectral radius of $ A_2 $ is only half as large as the spectral radius of $ A_1 $ for small step sizes $ h $, both methods require approximately the same number of iterations for convergence. Figure~\ref{fig:sr_psr} shows the dependence of the pseudospectral radii on the number of time steps for splitting $ \mc{P}_1 $. If the number of time steps is large, then the pseudospectral radii of the iteration matrices are almost equal. The pseudospectral radii were computed using Higham's \textit{Matrix Computation Toolbox}~\cite{MCT}. Here, the parameter $ \varepsilon $ for the computation of the $ \varepsilon $-pseudoeigenvalues was set to $ 10^{-3} $.

\begin{figure}[htb]
    \centering
    \begin{minipage}[c]{0.495\textwidth}
        \centering
        \subfiguretitle{a) $ \rho(A_1) $ and $ \rho_\varepsilon(A_1) $}
        \includegraphics[width=\textwidth]{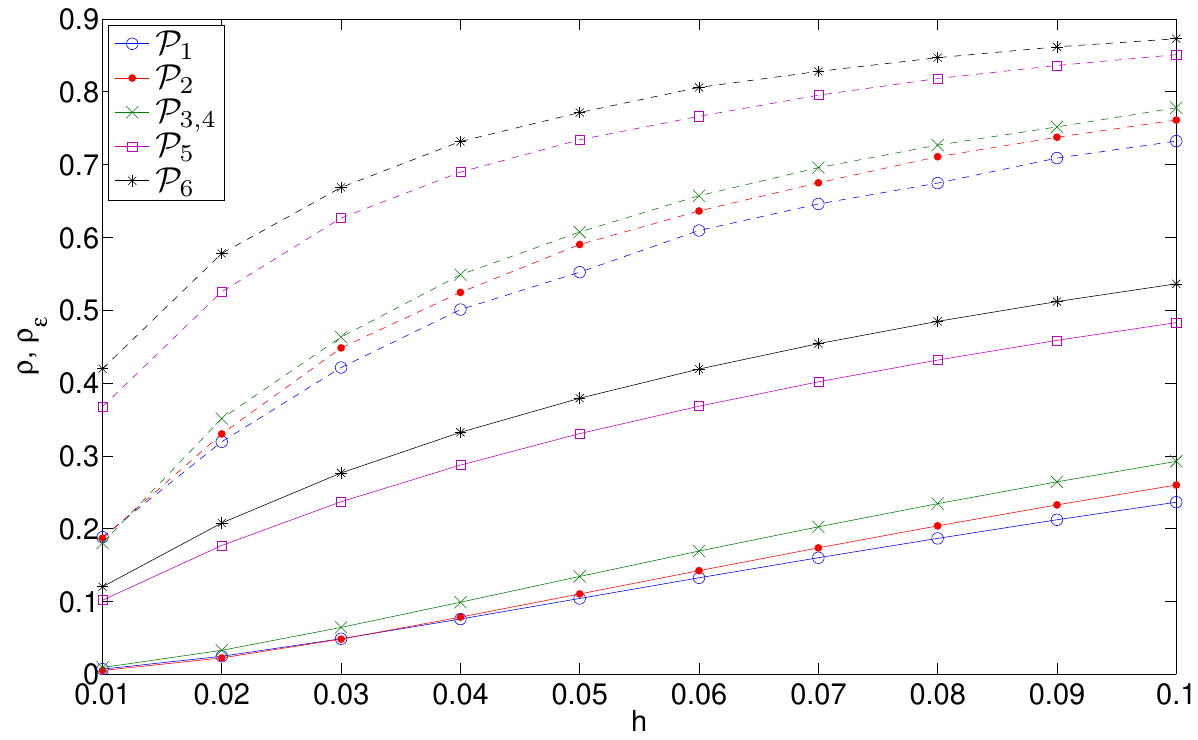}
    \end{minipage}
    \begin{minipage}[c]{0.495\textwidth}
        \centering
        \subfiguretitle{b) number of iterations $ k $}
        \includegraphics[width=\textwidth]{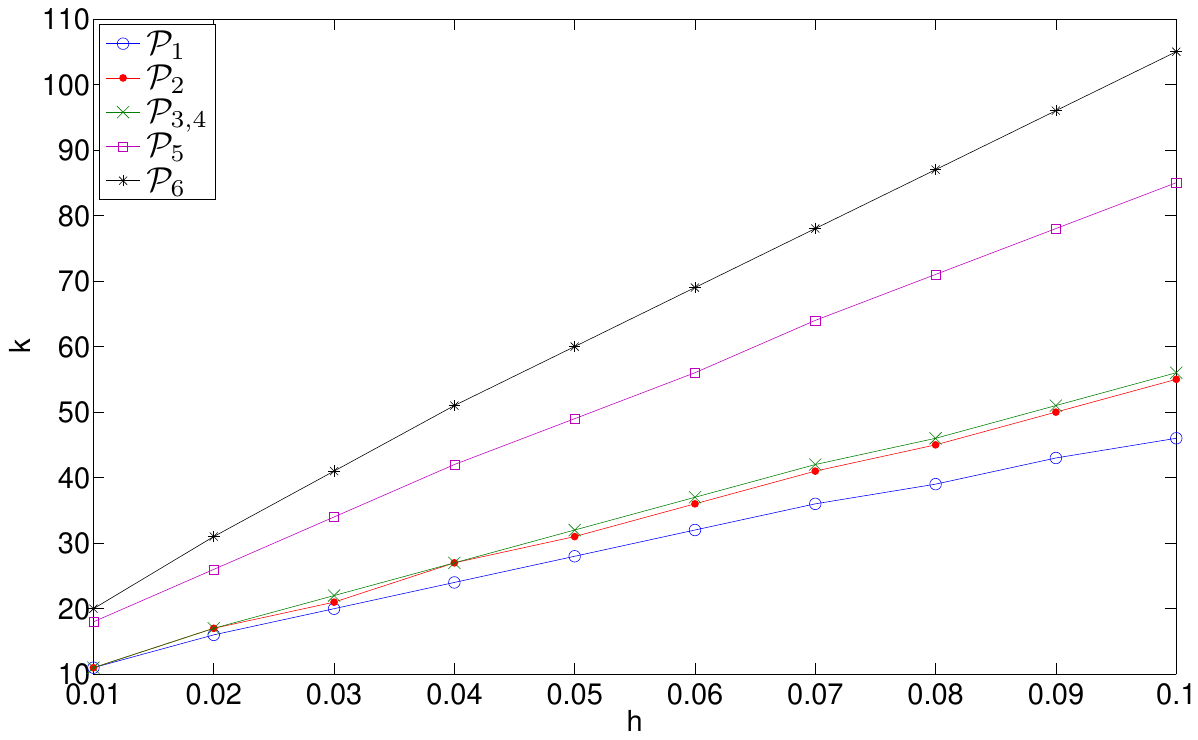}
    \end{minipage}
    \caption{Comparison of different splittings. a) Spectral radius (solid line) and pseudospectral
             radius (dashed line). b) Number of iterations required for convergence.}
    \label{fig:rho_k}
\end{figure}

Below, we compare $ \mc{P}_1 $ and $ \mc{P}_2 $ to the splittings generated by the heuristics described in Section~\ref{sec:Partitioning and Convergence}. The GSPN is irreducible and the spectral partitioning yields
\begin{equation}
    \mc{P}_3 =
        [1 \; 2 \; 3 \; 4 \; 5 \; 6 \; 8 \; 9 \; 10 \; 11 \; 12 \; 15 \; 16 \; 17 \; 18 \; 21 \mid
         7 \;  13 \;  14 \;  19 \; 20 \; 22 \; 23 \; 24]\,,
\end{equation}
while PARTY generates a balanced splitting
\begin{equation}
    \mc{P}_4 =
        [1 \; 2 \; 3 \; 4 \; 5 \; 6 \; 9 \; 10 \; 11 \; 15 \; 16 \; 18 \mid
         7 \; 8 \; 12 \; 13 \; 14 \; 17 \; 19 \; 20 \; 21 \; 22 \; 23 \; 24]\,.
\end{equation}
In Figure~\ref{fig:rho_k} the optimal splittings $ \mc{P}_1 $ and $ \mc{P}_2 $ are compared to the heuristic splittings $ \mc{P}_3 $ and $ \mc{P}_4 $. To illustrate the impact of this approach, two random splittings $ \mc{P}_5 $ and $ \mc{P}_6 $ are evaluated. Figure~\ref{fig:rho_k}a shows the spectral and the pseudospectral radii of the waveform relaxation operators using the implicit Euler method. The number of time steps was set to $ s = 50 $. Figure~\ref{fig:rho_k}b shows the number of iterations $ k $ required for convergence of standard waveform relaxation. Although the sizes of the parts of $ \mc{P}_3 $ and $ \mc{P}_4 $ are different, the results are virtually equivalent. Furthermore, the results are close to the results of the optimal splittings $ \mc{P}_1 $ and $ \mc{P}_2 $.

Now we combine both methods, the graph partitioning heuristics and adaptive waveform relaxation, and compare it to standard waveform relaxation. In addition, we subdivide the time interval $ [0, T] $ into $ 20 $, $ 25 $, and $ 30 $ windows of the same size and use standard waveform relaxation for each subinterval. We refer to these methods as FWR\subs{1}, FWR\subs{2}, and FWR\subs{3}, respectively. We set again $ T = 1 $ and $ \varepsilon = 10^{-4} $. Adaptive waveform relaxation generates---depending on the partitioning---between $ 24 $ and $ 27 $ windows. The runtime results are shown in Table~\ref{tab:Results_GSPN}. Note that the influence of the splitting on the convergence of adaptive waveform relaxation is much smaller than the influence on the standard waveform relaxation procedure.

\begin{table}[htb]
    \footnotesize
    \centering
    \caption{Runtime results for the GSPN in seconds.}
    \begin{tabular}{|l|rrrrrr|}
        \hline
           & \multicolumn{1}{|c}{$ \mc{P}_1 $} & \multicolumn{1}{c}{$ \mc{P}_2 $}
           & \multicolumn{1}{c}{$ \mc{P}_3 $} & \multicolumn{1}{c}{$ \mc{P}_4 $}
           & \multicolumn{1}{c}{$ \mc{P}_5 $} & \multicolumn{1}{c|}{$ \mc{P}_6 $} \\
        \hline
        AWR         & 0.104 & 0.103 & 0.103 & 0.104 & 0.105 & 0.106 \\
        FWR\subs{1} & 0.133 & 0.134 & 0.133 & 0.133 & 0.136 & 0.137 \\
        FWR\subs{2} & 0.121 & 0.126 & 0.125 & 0.125 & 0.130 & 0.132 \\
        FWR\subs{3} & 0.118 & 0.119 & 0.119 & 0.122 & 0.126 & 0.125 \\
        WR          & 0.619 & 0.622 & 0.621 & 0.622 & 0.869 & 1.071 \\
        \hline
    \end{tabular}
    \label{tab:Results_GSPN}
\end{table}

For this example, waveform relaxation using a fixed window size performs only slightly worse than adaptive waveform relaxation since the state probability distribution quickly converges to the equilibrium so that the extrapolation of the solution has almost no effect. However, the appropriate size of the windows is in general unknown prior to the simulation. Using adaptive waveform relaxation, the window sizes are generated and adjusted automatically.

To demonstrate the impact of the extrapolation and the adaptive windowing technique, we simulate 10 higher-dimensional networks $ Q_i $ with standard and adaptive waveform relaxation. For comparison, we subdivide the time interval into the same number of equally sized windows and use again standard waveform relaxation for each subinterval (FWR). The results are shown in Table~\ref{tab:Results}. We decompose each system into $ p = 2 \, n_\text{scc} $ blocks, with $ n_\text{scc} $ being the number of strongly connected components. The default partition is defined to be the balanced decomposition where the variables are assigned to the blocks without a previous permutation of the matrix.

\begin{table}[htb]
    \footnotesize
    \centering
    \caption{Runtime results for further examples in seconds.}
    \begin{tabular}{|l|r|r|rrr|rrr|}
        \hline
          & & & \multicolumn{3}{c|}{HVD+PARTY} & \multicolumn{3}{c|}{Default partition} \\ \cline{4-9}
          & \multicolumn{1}{c|}{$ n $} & \multicolumn{1}{c|}{$ n_\text{scc} $}
            & \multicolumn{1}{c}{AWR} & \multicolumn{1}{c}{FWR} & \multicolumn{1}{c|}{WR}
            & \multicolumn{1}{c}{AWR} & \multicolumn{1}{c}{FWR} & \multicolumn{1}{c|}{WR} \\
        \hline
        $ Q_1 $    &  100 &  1 &  0.45 &   0.88 &    5.89 &  0.53 &   0.95 &    8.14 \\
        $ Q_2 $    &  100 & 10 &  0.54 &   0.92 &    6.61 &  0.58 &   1.01 &   10.11 \\
        $ Q_3 $    &  200 &  1 &  0.98 &   2.01 &   12.01 &  1.04 &   3.01 &   18.81 \\
        $ Q_4 $    &  200 & 10 &  1.05 &   2.28 &   24.74 &  1.18 &   2.68 &   42.28 \\
        $ Q_5 $    &  400 &  1 &  8.77 &  20.16 &  204.25 &  9.42 &  21.02 &  251.74 \\
        $ Q_6 $    &  400 & 10 &  6.93 &  14.07 &   84.25 &  9.16 &  19.98 &  219.85 \\
        $ Q_7 $    &  800 &  1 & 27.13 &  69.18 &  346.62 & 36.21 &  74.27 &  589.85 \\
        $ Q_8 $    &  800 & 10 & 17.97 &  43.41 &  326.05 & 18.32 &  44.87 &  604.26 \\
        $ Q_9 $    & 1600 &  1 & 78.31 & 152.02 &  948.33 & 96.01 & 210.11 & 1550.62 \\
        $ Q_{10} $ & 1600 & 10 & 67.80 & 172.89 & 1434.59 & 73.06 & 203.06 & 2722.87 \\
        \hline
    \end{tabular}
    \label{tab:Results}
\end{table}

If the network consists of several strongly connected
components,then the horizontal-vertical decomposition is of big
importance for the convergence of waveform relaxation. To
illustrate the influence of the horizontal-vertical
decomposition, we simulate a $ 400 $ dimensional example which
consists of $ 20 $ strongly connected components. If we apply
PARTY directly to decompose the system into 40 subsystems, then
standard waveform relaxation takes approximately $ 110.93 $ sec
and adaptive waveform relaxation $ 6.77 $ sec. If we, on the
other hand, decompose the system first using the
horizontal-vertical decomposition and apply PARTY to the
individual strongly connected components, then the simulation
takes only $ 68.46 $ sec or $ 4.41 $ sec, respectively.

In summary, the combination of the horizontal-vertical decomposition and the different partitioning methods for undirected graphs enables a reliable and efficient splitting of the system for the subsequent standard or adaptive waveform relaxation.

\section{Conclusions}

The performance of waveform relaxation depends on many different influencing factors. One important criterion is the proper subdivision of the integration interval into smaller time windows. In this work, we proposed an adaptive waveform relaxation method which, depending on the previous time interval, generates appropriately sized time windows. In regions where the solution changes rapidly, small windows are computed and in regions where the solution changes slowly, large windows are computed. Decomposition of the system is also of great importance for the convergence of waveform relaxation. We analyzed the spectra and pseudospectra of the resulting waveform relaxation operators and introduced different graph partitioning heuristics in order to speed up the simulation. It was shown that it is possible to speed up the computation of high-dimensional differential equations using adaptive waveform relaxation along with appropriate partitioning heuristics.

\section{Acknowledgements}

The research in this document was performed in connection with contract W911QX-08-C-0069 with the U.S. Army Research Laboratory. The views and conclusions contained in this document are those of the authors and should not be interpreted as presenting the official policies or position, either expressed or implied, of the U.S. Army Research Laboratory or the U.S. Government unless so designated by other authorized documents. Citation of manufacturer's or trade names does not constitute an official endorsement or approval of the use thereof. The U.S. Government is authorized to reproduce and distribute reprints for Government purpose notwithstanding any copyright notation hereon.

The authors thank Alberto Sangiovanni-Vincentelli for valuable discussions and suggestions.

\bibliographystyle{elsarticle-num}
\bibliography{AWR}

\begin{thebibliography}{10}
\expandafter\ifx\csname url\endcsname\relax
  \def\url#1{\texttt{#1}}\fi
\expandafter\ifx\csname urlprefix\endcsname\relax\def\urlprefix{URL }\fi
\expandafter\ifx\csname href\endcsname\relax
  \def\href#1#2{#2} \def\path#1{#1}\fi

\bibitem{MetaNet}
H.~Jeong, B.~Tombor, R.~Albert, Z.~N. Oltvai, A.-L. Barab\'asi, The large-scale
  organization of metabolic networks, Nature 407 (2000) 651--654.

\bibitem{McMillan}
K.~L. McMillan, Symbolic Model Checking, 1st Edition, Kluwer Academic
  Publishers, 1993.

\bibitem{Berkooz}
P.~Holmes, J.~L. Lumley, G.~Berkooz, Turbulence, Coherent Structures, Dynamical
  Systems and Symmetry, 1st Edition, Cambridge University Press, 1996.

\bibitem{heart}
J.~L. Palladino, A.~Noordergraaf, Muscle contraction mechanics from
  ultrastructural dynamics, 1st Edition, Springer-Verlag, 1998.

\bibitem{icepic}
Z.~Ren, S.~B. Pope, A.~Vladimirsky, J.~M. Guckenheimer, The invariant
  constrained equilibrium edge preimage curve method for the dimension
  reduction of chemical kinetics, Journal of Chemical Physics 124 (2006)
  114111.

\bibitem{nevanlinear}
B.~Leimkuhler, U.~Miekkala, O.~Nevanlinna, Waveform relaxation for linear
  {RC}-circuits, Impact of Computing in Science and Engineering 3 (1991)
  123--145.

\bibitem{Modred}
P.~Benner, V.~Mehrmann, D.~C. Sorensen, Dimension Reduction of Large-Scale
  Systems, 1st Edition, Springer-Verlag, 2003.

\bibitem{Krylov}
E.~J. Grimme, Krylov projection methods for model reduction, Tech. rep. (1997).

\bibitem{Row05}
C.~W. Rowley, Model reduction for fluids, using balanced proper orthogonal
  decomposition, International Journal of Bifurcation and Chaos 15~(3) (2005)
  997--1013.

\bibitem{Alberto}
J.~K. White, A.~S. Sangiovanni-Vincentelli, Relaxation Techniques for the
  Simulation of VLSI circuits, 1st Edition, Kluwer Academic Publishers, 1987.

\bibitem{WaveformConv}
B.~Leimkuhler, Estimating waveform relaxation convergence, SIAM Journal on
  Scientific Computing 14 (1993) 872--889.

\bibitem{Guck}
J.~Guckenheimer, P.~Holmes, Nonlinear Oscillations, Dynamical Systems, and
  Bifurcations of Vector Fields, 1st Edition, Springer-Verlag, 1983.

\bibitem{Bur95}
K.~Burrage, Parallel and sequential methods for ordinary differential
  equations, Oxford University Press, 1995.

\bibitem{BS97}
M.~Bjorhus, A.~M. Stuart, Waveform relaxation as a dynamical system, in:
  Mathematics of Computation, Vol.~66, 1997, pp. 1101--1117.

\bibitem{Stoer}
J.~Stoer, R.~Bulirsch, R.~Bartels, W.~Gautschi, C.~Witzgall, Introduction to
  Numerical Analysis, 3rd Edition, Springer-Verlag, 2002.

\bibitem{LW97}
A.~Lumsdaine, D.~Wu, Spectra and pseudospectra of waveform relaxation
  operators, SIAM Journal on Scientific Computing 18 (1997) 286--304.

\bibitem{TE05}
L.~N. Trefethen, M.~Embree, Spectra and pseudospectra, Princeton University
  Press, 2005.

\bibitem{JO95}
Z.~Jackiewicz, B.~Owren, Convergence analysis of waveform relaxation methods
  using pseudospectra, Numerics {N}2-1995, Department of Mathematics,
  University of Trondheim (1995).

\bibitem{Hen03}
M.~R. Henzinger, Algorithmic challenges in web search engines, Internet
  Mathematics 1~(1) (2003) 115--126.

\bibitem{VKLM04}
S.~Varigonda, T.~Kalmar-Nagy, B.~LaBarre, I.~Mezic, Graph decomposition methods
  for uncertainty propagation in complex, nonlinear interconnected dynamical
  systems, in: 43rd IEEE Conference on Decision and Control, Vol.~2, 2004, pp.
  1794--1798.

\bibitem{Tutorial}
U.~von Luxburg, A tutorial on spectral clustering, Statistics and Computing
  17~(4) (2007) 395--416.

\bibitem{SSB09}
T.~Sahai, A.~Speranzon, A.~Banaszuk, Hearing the clusters in a graph: A
  distributed algorithm, CoRR abs/0911.4729.

\bibitem{Pre00}
R.~Preis, Analyses and design of efficient graph partitioning methods, Ph.D.
  thesis, University of Paderborn (2000).

\bibitem{GSPNBook}
M.~A. Marsan, G.~Balbo, G.~Conte, S.~Donatelli, G.~Franceschinis, Modelling
  with Generalized Stochastic Petri Nets, Wiley Series in Parallel Computing,
  John Wiley and Sons, 1995.

\bibitem{GSPNATM}
M.~A. Marsan, C.~Chiasserini, A.~Fumagalli, Performance models of handover
  protocols and buffering policies in mobile wireless {ATM} networks, IEEE
  Transactions on Vehicular Technology 50~(4) (2001) 925--941.

\bibitem{GSPNSoft}
C.~M. Woodside, Y.~Li, Performance petri net analysis of communications
  protocol software by delay-equivalent aggregation, in: Proc. 4th Int.
  Workshop on Petri Nets and Performance Models (PNPM'91), Melbourne,
  Australia, IEEE Comp. Soc. Press, 1991, pp. 64--73.

\bibitem{GSPNHard}
M.~A. Marsan, G.~Conte, A class of generalized stochastic petri nets for the
  performance evaluation of multiprocessor systems, ACM Transactions on
  Computer Systems 2 (1984) 93--122.

\bibitem{GSPNMaintenance}
M.~Molla-Hosseini, R.~M. Kerr, R.~B. Randall, R.~B. Platfoot, An inspection
  model with minimal and major maintenance for a flexible manufacturing cell
  using generalized stochastic petri nets, in: Proceeding of the 16th
  International Conference on Application and Theory of Petri Nets,
  Springer-Verlag, 1995, pp. 335--356.

\bibitem{GSPNequalCTMC}
M.~A. Marsan, G.~Conte, G.~Balbo, A class of generalized stochastic petri nets
  for the performance evaluation of multiprocessor systems, ACM Trans. Comput.
  Syst. 2~(2) (1984) 93--122.

\bibitem{GSPNLarge}
K.~L. McMillan, D.~K. Probst, A technique of state space search based on
  unfolding, in: Formal Methods in System Design, 1992, pp. 45--65.

\bibitem{MCT}
N.~J. Higham, The {Matrix Computation Toolbox},
  \verb|www.ma.man.ac.uk/~higham/mctoolbox|.

\end{thebibliography}

\end{document}